\documentstyle[aps,preprint]{revtex}

\begin{document}
\draft

\preprint{hep-ph@xxx/9507261 \qquad UAHEP954}
\title{SUSY Unification with Smooth Threshold Behavior\\}
\author{L. Clavelli and P.W. Coulter}
\address{
Department of Physics and Astronomy\\
University of Alabama\\
Tuscaloosa, Alabama 35487\\}
\maketitle
\begin{abstract}
          Going beyond the theta function approximation
          we discuss supersymmetric unification of gauge
          couplings with exact decoupling of light and
          heavy particles at energy scales below their masses.
          We find that the Minimal SUSY model is strongly
          disfavored while the Missing Doublet Model
          survives with GUT scale masses rising into the
          $10^{18}$ GeV region.
\end{abstract}
\pacs{11.30.Pb,14.80.Ly}

\newcommand{\be}{\begin{eqnarray}}
\newcommand{\ee}{\end{eqnarray}}
     Although there are tantalizing indications that the elementary
forces become approximately supersymmetric in the TeV region and grand
unified at a scale above $10^{16}$ GeV, there remain two possible problems
that have recently received some attention.  Problem number one is the
following.  The strong coupling constant at the $Z$ scale,
$\alpha_3(M_Z)$,
predicted in supersymmetric (SUSY) grand unified theory (GUT), while
agreeing with the apparent value at the $Z$ resonance, disagrees
markedly with the value expected from low energy measurements which
lies $12\%$ to $20\%$ lower ($3$ to $5~ \sigma$).  The difference
between the apparent value at the $Z$ and the actual value is one
of the possible indications of a light gluino\cite{CCFHPY} and can
be correlated with the excess in $Z \rightarrow b \overline{b}$
\cite{Rb,Erler,Roszkowski}.
Problem number two lies in the fact that the scale at which the
three gauge couplings appear to unify, $\sim 2 \cdot 10^{16}$ GeV,
lies about two orders of magnitude below what would be
expected in a more fundamental picture such as string theory.  It is
possible that the solution to both problems lies in physics at energy
scales far above that of present accelerators.  GUT scale effects are
at present highly model dependent.  One approach has been to assume
that, at the scale at which the $U(1)$ and $SU(2)$ couplings meet, the
$SU(3)$ coupling deviates from these by a fractional amount
$\epsilon$ as a parametrization of gravitational
or mass-splitting effects among the GUT scale particles.  In such an
approach the three couplings would not be seen to meet.  A five
percent effect at the GUT scale can lead to a ten percent difference
in the predicted $\alpha_3(M_Z)$ \cite{RUA,DMN,BMP}.
A suggested approach to the second problem
\cite{Ramond,Hempfling,Mohapatra}, is to seek an
intermediate scale at which new particles lie which could redirect the
gauge couplings to a unification nearer the Planck scale.
\par
     In order to explore alternative solutions to the above
problems, we have, in a recent paper\cite{cc2}, proposed the
following point of view.  We assume that gravitational effects
can be neglected.  With this assumption of negligible
gravitational effects, the model dependence of GUT scale effects
lies in the unknown degeneracy splitting among GUT scale particles as
well as the possibility of higher GUT scale Higgs representations.  In
the minimal supersymmetric model (MSSM), the GUT scale
masses are $M_V$ for the GUT scale gauge boson supermultiplet and
$M_\Sigma$ and $M_D$ for the GUT scale
Higgs supermultiplets.  If the theory is unified far above these
masses, it will remain unified (in the theta function approximation)
down to the maximum GUT scale mass below which calculable deviations
from unification will occur depending on the three GUT scale masses.
The full range of possible GUT scale effects in this model is then
determined by running over all possible values for $M_V$, $M_\Sigma$,
and $M_D$ consistent with proton decay and other experimental
constraints.  In the Missing Doublet Model (MDM)\cite{MDM} which has
a richer GUT scale Higgs spectrum, there are four masses to be varied,
$M_V$, $M_\Sigma$, $M_D$, and $M_\Phi$.  Above the highest GUT scale
particle, the MSSM is asymptotically free while the unified coupling
constant in the MDM grows with energy.  This leads to a possible new
constraint in the MDM, namely that the gauge coupling not become
grossly non-perturbative as one approaches the Planck scale.  It might
be considered an appealing feature of the MDM that the coupling constant
grows above the GUT scale becoming strong near the Planck scale where
unification with gravity might then be natural.  The result of this
exercise \cite{cc2} was that GUT scale degeneracy splitting
in the MSSM does not solve either of the two problems mentioned above.
On the other hand the MDM does lead to a value of $\alpha_3(M_Z)$
in good agreement with expectations from low energy data.  This latter
result is suggested in the earlier work of \cite{Hagiwara}
and is also noted in \cite{BMP}.  For
definiteness we adopt a low energy value for $\alpha_3(M_Z)$ spanning
the range between two recent low energy analyses\cite{CCY,Voloshin}.
Thus we take
\be
\alpha_3(M_Z) = \left\{ \begin{array}{cc} .104\pm.005 &\qquad
           \mbox{(heavy gluino case)}\\
 .119\pm.005 &\qquad \mbox{(light gluino case)} \end{array}
     \right.
\label{eq:low}
\ee
The lower extremes of this range correspond to the analysis of
\cite{CCY} while the upper extremes correspond to that of
\cite{Voloshin}.  Many other low energy analyses are consistent with
these ranges and virtually all (except for some from $\tau$ decay)
lie within two $\sigma$ which is
sufficient for our present considerations.  Another result
in the MDM was that the GUT scale, defined as the maximum
mass of GUT scale particles, can in fact increase into the $10^{17}$
GeV region\cite{cc2}.
\par
In the current work we wish to explore the further refinement of
smooth threshold behavior at the SUSY and GUT scales.  Recently
these smooth threshold effects have been given some renewed
attention \cite{BMP,Bastero}.  The first of these considers smooth
thresholds at the SUSY scale only, relying on an arbitrary shift of
$\alpha_3$ to parameterize GUT scale effects. The second does not
discuss the solution ranges for the GUT scale masses.
In the region of overlap our results are
consistent with theirs and we extend their conclusions.
\par
     It is well known that the variation in a gauge coupling
$\alpha(q)$ due to the one-loop propagator correction from a fermion
of mass $m_i$  in the Euclidean region is given\cite{GP} by
\be
4 \pi q {d \over{dq}} \alpha^{-1}(q) = - 2 b_i n_f(q/m_i)
\label{eq:rngrpf}
\ee
where
\be
  n_f(q/m) = 6 \int_{0}^{1} \frac{dx x^2(1-x)^2}{x(1-x)+m^2/q^2}
        = 1 - \frac{3}{2}(w^2-1) \left [1- \frac{w^2-1}{2w}\ln
        \left (\frac{w+1}{w-1} \right ) \right ]   \label{eq:nf}
\ee
with
\be
    w = \sqrt{1 + {4m^2/q^2}}   .  \label{eq:w}
\ee
For a bosonic loop one would have
\be
    4 \pi q {d \over{dq}} \alpha^{-1}(q) = - 2 b_i n_b(q/m_i)
    \label{eq:rngrpb}
\ee
with
\be
     n_b(q/m) = 1 + 3(w^2-1) \left [1 - \frac{w}{2} \ln \left (
             \frac{w+1}{w-1}\right )\right ].  \label{eq:nb}
\ee
$n_f(q/m)$ and $n_b(q/m)$ exhibit the decoupling behavior,
\be
  n_{f,b}(q/m) \rightarrow \left \{  \begin{array}{cc}
            1& \qquad q \gg m\\0& \qquad q \ll m \end{array}
            \right.
\label{eq:nflim}
\ee
So that, as an approximation, one can write for the contribution
from a particle of mass $m$ as
\be
  n_{f,b}(q/m) = \theta(q-m) \qquad \mbox{(theta function
  approximation)} . \label{eq:ntheta}
\ee
The $n$'s can be written analytically as perfect derivatives
\be
  n_{f,b}(q/m) = q {d \over {dq}} f_{f,b}(q/m)
\ee
with
\be
    f_f(q/m)=
                 \frac{w^2}{2} \left [ 1 - \frac{w^2-3}{2w}
                  \ln \left ( \frac{w+1}{w-1} \right ) \right ]
\label{eq:ff}
\ee
\be
  f_b(q/m) =
     {w^2 \over {2}} \left [ -1 + w \ln \left (\frac{w+1}{w-1}
     \right ) \right ]
\label{eq:fb}
\ee
Thus if one integrates eq.\ref{eq:rngrpf} from some
$q_0$ to $q_1$ the result
is
\be
   4 \pi \alpha^{-1}(q_0) = 4 \pi \alpha^{-1}(q_1)
      - 2 b \left [ f_f(q_0/m) - f_f(q_1/m) \right ]
\label{eq:alfinv}
\ee
with the corresponding result, changing $f_f$ for $f_b$,
in the case of a contribution from a boson.
If $q_0/m \ll 1 \ll q_1/m$ , this becomes
\begin{eqnarray}
  4 \pi \alpha^{-1}(q_0)& = &4 \pi \alpha^{-1}(q_1)
       - 2 b \left [ - \ln(q_1/m) + {5 \over 6}
       + {q_0^2 \over {10 m^2}} - \frac{5}{2} \frac{m^2}{q_1^2}
       \right.\nonumber\\&   & \mbox{} + {\cal O}\left.
       \left ( (q_0/m)^4, (m/q_1)^4 \ln(q_1/m) \right )\right ]
\label{eq:alfinvlim}
\end{eqnarray}
The theta function approximation is equivalent to keeping only the
first term in the square bracket.  The constant term can be taken
into account in the theta function approximation by imposing a
discrete shift (matching condition) in the couplings at
$q=m$ but in practice this shift is generally neglected.  If there are
particles in the vicinity of $q_0$ or $q_1$ the power series in
eq.\ref{eq:alfinvlim} is slowly convergent and the theta function
approximation becomes poor.  Of course, the theta function
approximation could be defined to be exact if the non-logarithmic
effects were properly incorporated elsewhere as in extracting
couplings from data\cite{CPP} but in practice this is not done and it
seems much more economical to include the threshold mass effects into
the running of the couplings.  Further discussion of the theoretical
basis for smooth decoupling ("Mass Dependent Subtraction Procedure")
is given in \cite{Bastero}. Practically all current grand unification
studies including \cite{Barger,Florida,Michigan,CCMNW,cc2}
have relied heavily on the theta function approximation to the
beta function.  A notable exception has been the work of \cite{Binoth}
where the full $n_f(q/m)$ was used for the top quark and gaugino
contributions together with the full $n_b(q/m)$ for the SUSY scalars.
However even in this work the effect of smooth thresholds at the GUT
scale was neglected.  In the current work we extend the smooth
threshold behavior to the GUT scale particles and to the low-lying
quarks, leptons, and gauge bosons.  Our purpose is to investigate the
differential effect on the gauge unification solutions when one goes
from theta function to smooth decoupling.  The effect on
the $b/\tau$ mass ratio, which is not part of the current study,
would be expected to be especially significant since the $b$ Yukawa
which, in the theta function approximation, is rising most rapidly
in the low energy region begins to be strongly suppressed as one
approaches the $b$ scale if one imposes a smooth decoupling.  The
GUT scale effects are also expected to be large.  Therefore, we do
not attempt to fit the top quark mass, $\tan \beta$ or the $b/\tau$
mass ratio.
\par
     In \cite{cc2} we considered, in the theta function
approximation, the effect of non-degeneracy among the GUT scale
particles in both the minimal supersymmetric model and the
missing doublet model.  In such a treatment the GUT scale is
considered to be the mass of the heaviest of the GUT scale particles
since if the couplings are unified there they remain unified at higher
energies.  With smooth decoupling the couplings can be assumed to be
unified far above the GUT scale masses but will begin to diverge as
one approaches the GUT scale.  We somewhat arbitrarily take unified
couplings at the Planck mass assuming that all GUT scale particles
have much smaller masses and that perturbation theory is still at
least qualitatively valid there.  Gravitational effects, which are
beyond the
scope of the present paper, are supplementary to the effects studied
here but should not void our qualitative conclusions.  In this paper
we restrict our interest to gauge coupling unification.  We
integrate the one-loop contributions exactly and analytically
including the full threshold behavior while treating the two loop
(including Yukawa) contributions, $\delta_i^{2L}$,
numerically with a crude (but
fully adequate) approximation to smooth decoupling ignoring the
two loop contributions of the GUT scale particles.  The two loop
contribution, $\delta_3^{2L}$ is found to be only 5 to 7 percent
of the analytic one loop contribution in the MDM and 9 to 17
percent in the MSSM and roughly linearly related to $\alpha_3(M_Z)$
in each case.  The b
coefficients are as given in \cite{Hagiwara,cc2} except that we
separate the GUT scale Higgs supermultiplet contributions into
separate contributions from bosons and fermions in the ratio of
1:2.  Similarly the contribution of the GUT scale gauge
supermultiplet separates into bosons and fermions in the ratio
11:(-2).  The top Yukawa, $\alpha_t(M_P)$, at the gauge
unification point is taken to be between $0.1$ and $0.9$.
Due to the fixed point behavior large values of the top Yukawa
rapidly evolve down to values of order $\alpha_3$ so that the
gauge couplings are relatively insensitive to the GUT scale
behavior and values of the Yukawa couplings.  Then
\be
  4 \pi \alpha_i^{-1}(M_Z) = 4 \pi \alpha_i^{-1}(M_P)
  - 2 \sum_{j} b_i(j) \left [f_j(M_Z,m_j)-f_j(M_P,m_j) \right ]
                    + \delta_i^{2L}  .
\label{eq:alfinvmz}
\ee
At the Planck mass, $1.22\cdot 10^{19}$ GeV, we take the unification
condition
\be
 4 \pi \alpha_3^{-1}(M_P)-1 = 4 \pi \alpha_2^{-1}(M_P) - {2 \over 3}
     = 4 \pi \alpha_1^{-1}(M_P) \equiv 4 \pi \alpha_0^{-1}  .
\label{eq:alf3invmp}
\ee
We choose the Planck scale gauge coupling, $\alpha_0$, and the Planck
scale top Yukawa, $\alpha_t(M_P)$, at random as we do for the various
GUT scale masses, $M_V, M_D, M_\Sigma, M_\Phi$, and the SUSY scales,
$m_0$, and $m_{1/2}$. For perturbative consistency, however, we
require that $\alpha_0 < 1/4$, $\alpha_t(M_P)<1$, and $m_j/M_P <1/5$.
The SUSY masses are chosen with
$100$ GeV $<m_0 <1$ TeV and $50$ GeV $<m_{1/2}< 330$ GeV.  Splitting
between partners of left and right handed fermions is neglected.
The squark and slepton masses are defined in terms of $m_0$ and
$m_{1/2}$ as in \cite{Bastero}.  That is: $m_{\tilde q}^2 = m_0^2
+7m_{1/2}^2$ , $m_{l,L}^2 = m_0^2 + 0.5 m_{1/2}^2$, and $m_{l,R}^2
= m_0^2 + 0.15 m_{1/2}^2$.  We
discard as non-solutions values of these parameters inconsistent with
the renormalization group running and the experimental values
$\alpha^{-1}(M_Z)=127.9\pm.2$ and $\sin^2\theta_W=.2320\pm.0008$.
Further technical details on our ``top-down'' approach
may be found in \cite{cc2}.  By such Monte Carlo methods it is
possible to completely determine the multidimensional solution space.
\par
In \cite{cc2} it was found that the MSSM, with or without GUT scale
degeneracy breaking, was inconsistent with the low energy data on
$\alpha_3$ if
proton decay constraints and a theoretically desirable SUSY scale
below $1$ TeV were imposed.  This was also noted in the degenerate
case by \cite{Hagelin} and has been emphasized more recently by
\cite{Roszkowski}.  The lower limits on $\alpha_3(M_Z)$
in the MSSM with theta function decoupling are consistent with those
found by other authors \cite{LP,Wright}.
If one requires only agreement with the LEP values of $\alpha_3$ the
inconsistency is not apparent.  One of the conclusions of \cite{cc2}
was that this inconsistency disappears if the MSSM is replaced by the
missing doublet model.  This result was also noted in \cite{BMP}.
These considerations are independent of
whether or not the gluino is light (in the GeV region) as is still not
experimentally ruled out.  In table $1$ we compare the smooth threshold
results for $\alpha_3(M_Z)$ and the GUT scale masses with the
results of \cite{cc2} where GUT scale degeneracy was broken but
sharp (theta function) decoupling was used.  Slight differences
between our current requirements and those of \cite{cc2} with regard
to $\sin^2\theta_W$ and the $b/\tau$ mass ratio do not
affect the clear, qualitative conclusions that can be drawn from the
comparison in table $1$.  The numbers in table $1$ are shown in the
heavy gluino scenario but are not sensitive to this choice.
For example, with smooth decoupling, if one puts $m_{1/2}=0$ (light
gluino option) the minimum $\alpha_3(M_Z)$ drops only to $0.113$ in
the MDM and only to $0.169$ in the MSSM.  On the other hand
Ref.\cite{BMP} finds some preference in the MSSM for $m_{1/2} \ll m_0$
implying at least a relatively light gluino.  However, our results
indicate that with smooth decoupling at the GUT scale the MSSM cannot
be saved by this mechanism. In the MDM we discard solutions with
$\alpha_3(M_Z)$ above $0.135$ since these seem of no
phenomenological interest.  The same requirement in the MSSM would
eliminate all solutions leading to our conclusion that the MSSM is
no longer viable when smooth decoupling is taken into account.
\par
To summarize the conclusions of this study we may say the following.
In \cite{cc2} we noted that the MSSM with theta function threshold
behavior predicted an $\alpha_3(M_Z)$ inconsistent with
extrapolations from low energy data.  Our current results strongly
reinforce this conclusion and
disfavor the MSSM even if the higher LEP values of $\alpha_3(M_Z)$
are used.  Because of the very large values of $\alpha_3$ predicted
in the MSSM with smooth thresholds, current estimates of
gravitational effects cannot salvage the situation without
calling into question the successful prediction of $\sin^{2}\theta_W$.
For this reason it is our opinion that the MSSM is highly unlikely to
be realized in nature.  The low energy measurements of $\alpha_3$
and $\sin^2\theta_W$, therefore, strongly suggest a richer GUT scale
Higgs structure such as that given in the MDM.  The phenomenological
superiority with respect to grand unification of the MDM over the
MSSM was first pointed out in \cite{cc2}.  This model, when smooth
threshold behavior is taken into account, also contains unification
solutions with the heaviest GUT scale particles in the
$10^{18}$  GeV region as suggested by string theory.
If we require $\alpha_3(M_Z)<0.12$ we find, in fact, that all the
solutions in the MDM have $M_\Phi> 5\cdot 10^{17}$ GeV.
All the MDM solutions have the leptoquark gauge boson
supermultiplet in the $10^{16}$ GeV region or below
suggesting that proton decay could be dominated by the ep decay
modes expected in non-supersymmetric SU(5).  If we compare the
unification lower limits on $\alpha_3(M_Z)$ from table 1 with the
results from low energy analyses given in eq.\ref{eq:low}
we see that the light gluino option is somewhat favored.  However,
if there are $10\%$ effects from gravity
or other sources this preference might be eliminated.
\par
The authors acknowledge useful comments on this work from
Professor P.H. Cox of Texas A\&M University-Kingsville.
This work was supported in part by the Department of Energy
under grant DE-FG05-84ER40141.

\widetext
\vspace{5mm}
\begin{table}
\caption{
 Minimum and maximum values of $\alpha_3(M_Z)$ and the GUT
 scale masses in the unification solution space of the MSSM and MDM
 with either sharp or smooth decoupling.
 Underlined values are upper or lower limits imposed for
 phenomenological reasons discussed in the text.}
\vspace {3mm}
\tabskip=0pt\offinterlineskip
\baselineskip=18pt
\def\tablerule{\noalign{\hrule}}
 \halign to 6.5truein{\vrule#\tabskip=1em plus2em
 &#\hfil\strut&\hfill#\hfill&\hfill#\hfill&\hfill#\hfill&
    \hfill#\hfill&\vrule#
    \hfil\tabskip=0pt\cr
 \tablerule
 height4pt&\omit&\omit&\omit&\omit&\omit&\cr
 &\omit&MSSM&MSSM&MDM&MDM&\cr
 height2pt&\omit&\omit&\omit&\omit&\omit&\cr
 &\omit&sharp&smooth&sharp&smooth&\cr
 height4pt&\omit&\omit&\omit&\omit&\omit&\cr
 &$\alpha_3(M_Z)$&$(.117,.133)$&$(.174,.24)$&$(.095,.114)$&
       $(.116,\underline{.135})$&\cr
 &$M_V$ (GeV)&$(0.7,82)10^{15}$&$(1.8,130)10^{15}$&$(2.3,21)10^{15}$
       &$(\underline{1.0},6.4)10^{15}$&\cr
 &$M_D$ (GeV)&$(\underline{1.0},24)10^{16}$&$(\underline{.01},
    \underline {2.4})10^{18}$&$(\underline{1.0},18)10^{16}$
       &$(\underline{1.0},24)10^{16}$&\cr
 &$M_\Sigma$ (GeV)&$(0.3,15)10^{16}$&$(\underline{.01},
       \underline{2.4})10^{18}$&$(\underline{1.2},6.7)10^{16}$
       &$(.021,\underline{2.4})10^{18}$&\cr
 &$M_\Phi$ (GeV)&-&-&$(\underline{1.0},7.8)10^{16}$
       &$(0.21,\underline {2.4})10^{18}$&\cr
 height4pt&\omit&\omit&\omit&\omit&\omit&\cr
 \tablerule  } 
\end{table}

\begin{references}
\bibitem{CCFHPY}L. Clavelli, P.W. Coulter, B. Fenyi, A. Hester,
 P. Povinec, and K. Yuan, Phys. Lett. B291, 426 (1992)
\bibitem{Rb}L. Clavelli, hep-ph/9410343, Mod. Phys. Lett.
   {\bf A10}, 949 (1995)
\bibitem{Erler}J. Erler and P. Langacker, hep-ph/9411203
\bibitem{Roszkowski}L. Roszkowski and M. Shifman, hep-ph/9503358
\bibitem{RUA}D. Ring, S. Urano, and R. Arnowitt, hep-ph/9501247
\bibitem{DMN}T. Dasgupta, P. Mamales, and P. Nath, hep-ph/9501325
\bibitem{BMP}J. Bagger, K. Matchev, and D. Pierce, hep-ph/9501277
\bibitem{Ramond}S.P. Martin and P. Ramond, Phys. Rev. {\bf D51},
   6515 (1995)
\bibitem{Hempfling} R. Hempfling, hep-ph/9502201
\bibitem{Mohapatra}B. Brahmachari and R. Mohapatra, hep-ph/9505347
\bibitem{cc2}L. Clavelli and P.W. Coulter, Phys. Rev.
   {\bf D51}, 3913 (1995)
\bibitem{MDM}B. Grinstein, Nucl. Phys. {\bf B206}, 387 (1982);
  A. Masiero, D.V. Nanopoulos, K. Tamvakis, and T. Yanagida,
  Phys. Lett. {\bf 115B}, 380 (1982)
\bibitem{Hagiwara}K. Hagiwara and Y. Yamada,
     Phys. Rev. Lett. {\bf 70},709 (1993)\\
     Y. Yamada, Zeit. f. Phys. {\bf C60}, 83 (1993)
\bibitem{CCY}L. Clavelli, P.W. Coulter, and K. Yuan, Phys. Rev.
     {\bf D47},1973, 1993; \\L. Clavelli and P.W. Coulter,
     Phys. Rev. {\bf D51}, 1117 (1995)
\bibitem{Voloshin}M. Voloshin, hep-ph/9502224, IJMPA in press
\bibitem{Bastero}M. Bastero-Gil and J. Perez-Mercader, hep-ph/9506222
\bibitem{GP} H. Georgi and D. Politzer, Phys. Rev.
     {\bf D14}, 1829  (1976)
\bibitem{CPP}P. Chankowski, Z. Pluciennik, and S. Pokorski,
     Nucl. Phys. {\bf B439}, 23 (1995)
\bibitem{Barger}V. Barger, M.S. Berger, and P. Ohnmann, Phys. Rev.
     {\bf D47}, 1093 (1993)
\bibitem{Florida}D. Castano, E.J. Piard, and P. Ramond, Phys. Rev.
     {\bf D49}, 4882 (1994)
\bibitem{Michigan}G.L. Kane, C. Kolda, L. Roszkowski, and J. Wells,
     Phys. Rev. {\bf D49}, 6173 (1994)
\bibitem{CCMNW}M. Carena, L. Clavelli, D. Metalliotakis, H.-P. Nilles,
     and C. Wagner, Phys. Lett. {\bf B317}, 346 (1993)
\bibitem{Wright}B. Wright, hep-ph/9404217
\bibitem{Binoth} T. Binoth and J.J. van der Bij, Z. Phys.
     {\bf C58}, 581 (1993)
\bibitem{Hagelin} J. Hagelin, S. Kelley, and V. Ziegler,
     Phys. Lett. {\bf B342}, 145 (1995)
\bibitem{LP}P. Langacker and N. Polonsky, Phys. Rev. {\bf D49},
    1454 (1994)
\end{references}
\end{document}